**Title: "The Integration of Artificial Intelligence in Undergraduate Medical Education in Spain: Descriptive Analysis and International Perspectives"**

**Brief title:** AI in Spanish Medical Education

**Authors**: Ana Enériz Janeiro MSc[*1], Karina Pitombeira Pereira PhD [*1], Julio Mayol PhD [2, 8], Javier Crespo PhD [3, 8], Fernando Carballo PhD [4,8], Juan B. Cabello PhD [5,8], Manel Ramos-Casals PhD[6], Bibiana Pérez Corbacho MSc[1], Juan Turnes PhD [**1, 7, 8]

* These authors contributed equally to this work
** Corresponding Author:
    Email: jturnesv@gmail.com
    Phone: +34 986 800 907

**Affiliations:**

([1]) IDARA Research Group, Galicia Sur Health Research Institute (IISGS), Vigo, Spain.

([2]) Clinical Hospital San Carlos, San Carlos Health Research Institute (IdISSC), Complutense University of Madrid Medical School, Madrid, Spain

([3]) Department of Medicine and Psychiatry, Faculty of Medicine. Clinical and Translational Research Group in Digestive Diseases. Valdecilla Research Institute (IDIVAL), Santander, Spain

([4]) University of Murcia Murcia, Spain

([5]) Critical Appraisal Skills Programme: CASP Spain.

([6]) Postgraduate Degree Program in Artificial Intelligence in Medicine, Department of Medicine, University of Barcelona, ICMiD, Hospital Clínic, Barcelona, Spain

([7]) Department of Gastroenterology and Hepatology, Pontevedra University Hospital Complex, Pontevedra, Spain

([8]) MedicineAI, Madrid, Spain

**Abstract**

Background: AI is transforming medical practice and redefining the competencies that future healthcare professionals need to master. Despite international recommendations, the integration of AI into Medicine curricula in Spain had not been systematically evaluated until now.

Methods: A cross-sectional study (July–September 2025) including Spanish universities offering the official degree in Medicine, according to the "Register of Universities, Centers and Degrees (Registro de Universidades, Centros y Títulos - RUCT)". Curricula and publicly available institutional documentation were reviewed to identify courses and competencies related to AI in the 2025–2026 academic year. The analysis was performed using descriptive statistics.

Findings: Of the 52 universities analyzed, ten (19·2%) offer specific AI courses, whereas 36 (69·2%) include no related content. Most of the identified courses are elective, with a credit load ranging from three to six ECTS, representing on average 1·17% of the total 360 credits of the degree. The university of Jaén is the only institution offering a compulsory course with AI content. The territorial analysis reveals marked disparities: Andalusia leads with 55·5% of its universities incorporating AI training, while several communities lack any initiative in this area.

Interpretation: The integration of AI into the medical degree in Spain is incipient, fragmented, and uneven, with a low weight in ECTS. The limited training load and predominance of elective courses restrict the preparation of future physicians to practice in a healthcare environment increasingly mediated by AI. The findings support the establishment of minimum standards and national monitoring of indicators

**Funding:** No funding was received for this study.

**Keywords:** artificial intelligence, generative artificial intelligence, medical education, curriculum, Spain

**Research in context panel**

Evidence before this study:

We searched PubMed, Scopus, and Web of Science up to July 2025 for studies on the inclusion of AI courses or formal AI training in undergraduate medical curricula. We found international initiatives describing the introduction of AI in medical education, mainly in North America and Asia, but these were often limited to individual universities or pilot projects. To our knowledge, no nationwide systematic evaluation of AI integration into undergraduate medical curricula in Spain had been reported.

Added value of this study:

This study provides the first national overview of how AI is being integrated into undergraduate medical education in Spain. By analyzing the official curricula of all 52 universities offering Medicine, we demonstrate that AI training is scarce, mostly elective, and carries minimal academic weight. We also identify striking regional disparities: while some regions such as Andalusia have implemented courses, others have none. This comprehensive analysis highlights the fragmented and uneven adoption of AI in medical curricula.

Implications of all the available evidence:

Our findings show that AI education in Spanish medical schools is at an early and fragmented stage, with low ECTS weight and a predominance of elective courses. This limited approach risks leaving future physicians underprepared for a healthcare environment increasingly shaped by AI. Considering international recommendations, the evidence supports the establishment of minimum standards, stronger national coordination, and systematic monitoring of AI training in medical curricula to ensure equitable and consistent preparation across regions.

# 1. Introduction

Artificial intelligence (AI) has emerged as a transformative technology in the field of healthcare, revolutionizing everything from medical diagnosis to hospital management and biomedical research.[1] This technological revolution is not only transforming current clinical practice but also redefining the competencies that future healthcare professionals will need to practice medicine effectively in the coming years.[2]

The integration of artificial intelligence into medical curricula is globally recognized as a priority, underscoring the need to prepare future physicians with the technical and ethical competencies required to ensure the responsible use of these technologies. Moreover, there is an urgent need to develop structured curricular frameworks that systematically incorporate AI education into medical schools.[3]

However, the translation of these recommendations into concrete curricula has been uneven and fragmented. The advent of generative AI adds urgency, as it amplifies both the educational potential (e.g., support for writing and supervised reasoning) and the risks (hallucinations, data leakage, authorship attribution), making it essential to teach strategies for verification, documentation, and boundaries of use. The lack of AI training is also acknowledged by the students themselves. Global and regional surveys show broad support for its teaching, calling for its inclusion in curricula, while also expressing concern about ethical issues, the doctor–patient relationship, and its impact on specialty choice.[4,5] In a study conducted with medical students in Spain, 83% considered it essential to acquire knowledge of artificial intelligence for their future professional practice.[6]

To address this need, several countries and organizations have begun implementing systematic initiatives to incorporate AI into medical education. In Canada, curricular proposals and needs assessments have been developed to identify key competencies in areas such as ethics, legal aspects, clinical application, and collaborative work.[7] In the United States, several medical schools have advanced in the formal integration of AI into their curricula, with prominent examples including Harvard Medical School and the Icahn School of Medicine at Mount Sinai, the latter being a pioneer in its cross-curricular incorporation.[8] Complementarily, international studies involving medical students and faculty consistently indicate that AI training remains limited and urgently call for curricular updates to address the digital transformation in healthcare.[4,9,10]

The Spanish context presents relevant particularities: 52 medical schools, a 360-ECTS (European Credit Transfer and Accumulation System) degree program, and quality assurance frameworks that both enable and require transparent curricular structures. Despite the proliferation of isolated initiatives, to the best of our knowledge there is no national, systematic, and reproducible characterization of how AI is being incorporated into curricula, what educational weight it receives (in ECTS), whether it is compulsory or elective, and to what extent it includes specific content on generative AI.

The primary objective of this study is to provide the first analysis of the current state of AI integration in undergraduate medical education programs in Spain, based on reproducible criteria and supported by an open dataset for verification and reuse. The results will allow us to characterize the presence and nature of AI teaching in Spanish universities offering a medical degree, assess differences between public and private institutions, and provide valuable information for educational policymakers, universities, and medical educators.

## 2. Methods

A cross-sectional study was conducted between July and September 2025 to analyze the integration of AI education in undergraduate medical programs in Spain. A census of all Spanish universities offering an official degree in Medicine was performed according to the official website of the "Ministry of Science, Innovation and Universities (Ministerio de Ciencia, Innovación y Universidades)", using the "Register of Universities, Centers and Degrees (Registro de Universidades, Centros y Títulos)" (RUCT) [11], following the search parameters specified in Table 1.

This registry includes 33 universities offering a degree in Medicine in Spain, updated to 2021; therefore, this consultation was complemented with an exhaustive search of universities offering the degree in the 2025–2026 academic year (last verified on September 4, 2025).

2.1. Search Strategy

The search strategy was based on the selection of institutions currently offering an active degree in Medicine. The institutional websites of each university were reviewed to identify information on curricula, courses, and AI-related competencies within the Medicine degree for the 2025–2026 academic year. This study relied exclusively on publicly available information from official websites and public curricular documents, including study plans, course programs, and teaching guides.

A university is considered to offer a specific AI course if the Spanish term "Artificial Intelligence (Inteligencia Artificial)" or "AI (IA)" appears in the course title and/or if more than 50% of the content is dedicated to artificial intelligence. Within this group, courses are classified as generative AI courses when the Spanish term "generative AI (IA generativa)" is specified in the course title, syllabus, or teaching guide.

The existence of courses that might include content on artificial intelligence, even if not explicitly stated in the title, was also examined. The Spanish-keywords used for this selection were: "new technologies (nuevas tecnologías)", "informatics (informática)", "communication (comunicación)", "digital media (medios digitales)" or "digital health (salud digital)". Thus, a university is considered to offer a course "similar to AI or appearing in competencies" when artificial intelligence accounts for less than 50% of the course syllabus and/or when reference is made to the keywords.

Universities are considered to have no AI courses when they do not include any course on AI or similar to AI.

For each course, the following information was collected: university, region, type (public/private), title, status (compulsory/elective), training credits, academic year in which it is offered, department (or, alternatively, the specialty of the faculty teaching it), and main course content.

No personally identifiable information was collected, and no surveys were conducted with individuals. Therefore, ethics committee approval was not required according to Spanish regulations for research using public data.

To ensure the quality and accuracy of the collected data, each data entry was independently verified by two researchers and validated by an external evaluator not involved in the data collection process.

**Statistical Analysis**

A descriptive analysis of the data was performed using Python 3.11 with the pandas, matplotlib, and seaborn libraries. Absolute and relative frequencies of the main variables were calculated to identify the proportion of Artificial Intelligence (AI) courses included in the Medicine curricula in Spain, as well as the presence of courses related to AI, defined as "AI-similar" courses. Likewise, a territorial analysis was carried out, differentiating the AI training offer by Autonomous Region, to contextualize the provision of AI training within the Spanish educational system

**3. Results**

3.1. Scope and Distribution

The analysis included a total of 52 Spanish universities offering a degree in Medicine, representing all institutions with officially recognized programs in Spain for the 2025–2026 academic year. Of these universities, 36 (69·2%) were public institutions and 16 (30·8%) were private institutions.

The territorial distribution shows a concentration of medical schools in certain regions. Madrid leads with ten universities (19·2% of the total), followed by Andalusia with nine (17·3%) and Catalonia with eight (15·4%). The Valencian Community has five universities (9·6%), while the Canary Islands have three institutions (5·8%). The remaining autonomous communities each have between one and two universities, revealing a heterogeneous distribution that reflects both population density and the historical development of medical education in each territory.

3.2. Distribution of AI Education

Of the 52 universities analyzed, ten (19·2%) offer courses specifically dedicated to artificial intelligence in medicine (Table 2). Another six universities (11·5%) include AI-related content within broader courses or mention AI-related competencies in their learning objectives (Table 3). Therefore, considering both universities with specific AI courses and those with similar content, a total of 16 institutions (30·8%) have incorporated Artificial Intelligence into medical education in some form, whereas 36 institutions (69·2%) have not yet included explicit AI content in their Medicine curricula.

By type of institution, among the 36 public universities, seven (19·4%) offer this type of training, whereas among the 16 private universities, three (18·8%) include AI courses in their curricula (Table 4).

The territorial analysis reveals a significant disparity among autonomous communities. Andalusia leads, with five of its nine universities (55·5%) offering AI training. It is followed by Catalonia, with two of eight universities (25%), and Madrid, with two of ten universities (20%). In the Canary Islands, one of the three universities (33·3%) has incorporated this training. By contrast, regions such as Aragon, Balearic Islands, Cantabria, Castile-La Mancha, Castile and León, Valencian Community, Galicia, Murcia, Navarra and the Basque Country do not have universities that have integrated AI courses into their Medicine curricula (Figure 1).

3.3. Characteristics of AI Education

Among the universities that include specific courses on artificial intelligence in their curricula, most of them are offered as electives. The only exceptions are the University of Jaén, which offers a compulsory course, and University of Córdoba, which includes a core course (Table 2). The course at University of Jaén represents a unique case, as it is the only institution that offers a compulsory course with artificial intelligence content within the Medicine degree. Although the course does not explicitly mention AI in its title, "New Clinical and Biomedical Information Technologies (Nuevas Tecnologías de la Información Clínica y Biomédica)", its syllabus explicitly incorporates competencies related to this field, making it the only identified case of mandatory curricular integration.

3.4. ECTS credits

The analysis of the credit load dedicated to Artificial Intelligence in the Medicine degree reveals that the AI specific courses identified range between three and six ECTS credits (Table 2). In relation to the total of 360 credits required for the degree, AI training constitutes, on average, 1·17% of the medical curriculum. Nevertheless, since most of these courses are offered as electives, they are not included in the compulsory credit load, and not all students enroll in them. Consequently, the actual weight of AI training within the degree program may be regarded as even smaller. Complutense University of Madrid shows the highest dedication with six credits, while other institutions generally offer three-credit courses. In some cases, AI content is included within broader courses on medical informatics or health technologies, which makes it difficult to quantify its exact weight.

**4. Discussion**

Despite progress in the incorporation of artificial intelligence in the clinical and medical fields, its presence in medical curricula in Spain remains very limited and heterogeneous. Among the institutions that have included this course, most do so through elective courses, indicating that its presence in the curriculum is still marginal and incipient (1·17% of the 360 ECTS). Moreover, there is considerable heterogeneity in the approaches adopted, ranging from specific courses to the inclusion of AI content within transversal competencies. This reflects the absence of a common framework and clearly defined learning objectives for its integration into medical education. The findings of this study are consistent with recent research [12,13], This highlights a global gap in preparing future physicians for the digital era. These disparities raise concerns about the readiness of future health professionals and the ability of different healthcare systems to fully benefit from AI technologies. These findings reinforce the need for a national framework of core competencies in medical AI, like those promoted by international bodies such as the AAMC (Association of American Medical Colleges) and the GMC (General Medical Council), which would help establish common standards and reduce fragmentation .[14,15]

An international study conducted across 192 faculties of Medicine, Dentistry, and Veterinary Medicine in various countries examined the integration of Artificial Intelligence into curricula and students' perceptions of its impact on clinical practice. The Spanish subset of data (74 students from five universities) indicated that their programs lacked AI courses and showed strong interest in receiving additional training. Furthermore, 74% reported limited knowledge, and 80% expressed a sense of unpreparedness to apply AI in their future professional practice.[13] This gap between high student demand and the limited educational offering necessitates coordinated policy actions, including the transition of elective courses into core or compulsory content and the establishment of dedicated faculty training programs in AI.[16]

This gap between student demand and educational provision becomes even more evident when regional differences are considered. The territorial analysis reveals a marked disparity in the integration of AI into Medicine curricula: while Andalusia leads in AI incorporation, ten communities have no universities offering AI training during the medical degree. This pattern is consistent with findings from international studies, which also report uneven implementation across regions and countries.[4] These findings highlight the urgent need to design national strategies that prevent fragmented training and ensure equitable access to AI competencies among future healthcare professionals.

The United States has undergone a remarkable transformation in recent years, shifting from concerns about students' use of AI to the active teaching of these technologies. The Icahn School of Medicine at Mount Sinai became the first U.S. institution to fully incorporate AI into its medical education program, integrating tools such as ChatGPT into multiple aspects of the curriculum.[17] Since 2018, the Stanford Center for Artificial Intelligence in Medicine & Imaging (AIMI) has been promoting a formal curriculum in clinical AI for medical students, which includes fundamentals, applications, and ethical and legal aspects, with clear guidelines for its integration into medical training. [18]

Europe presents a mixed landscape, generally more advanced than Spain. Countries such as Germany, Belgium, and the United Kingdom have launched pilot programs to assess essential AI competencies and to develop curricular frameworks adapted to their national contexts.[19–21]

In Spain, the need to address this knowledge gap is particularly urgent given the aging population, the growing prevalence of chronic diseases, and the pressing need to optimize healthcare resources.[22] The integration of AI into medicine offers potential tools to improve diagnostic accuracy, personalize treatments, optimize workflows, and reduce clinical errors.[23] However, it also poses significant challenges related to ethics, professional responsibility, the interpretation of algorithmic results, and the need to keep human clinical judgment at the center of medical care.[24] Only if healthcare professionals are adequately prepared to implement AI effectively and ethically can the current challenges of the healthcare system be addressed. These inequalities are not merely educational but have public health and international equity implications, as insufficient AI training could deepen gaps in the quality of care and limit the adoption of innovations in the Spanish healthcare system compared with other European countries.[13]

A second element of "urgency" is that the emergence of generative AI is progressing exponentially and is already transforming medical practice: models such as MedPaLM 2 and MedLM are already in use at institutions like HCA Healthcare and BenchSci to generate clinical notes, answer complex medical queries, and automate routine tasks.[16,25] This evolution is expected to accelerate in the short term, as multimodal models such as MedGemini are already achieving 91.1% accuracy in combined medical tasks based on text and images, demonstrating their immediate relevance in clinical settings.[26] It is therefore urgent to adapt medical education in this field, as the magnitude of the change will be measured in years rather than decades. Moreover, delays in the curricular integration of AI may carry significant economic and opportunity costs, both for universities and for the national health system, by limiting the ability to train professionals capable of leading digital innovation projects and reducing inefficiencies in clinical practice.[1,23] In this context, it is crucial that curricula incorporate explicit strategies for the supervision and responsible use of generative AI, including guidelines for verification, documentation, and boundaries of use, as highlighted in recent educational proposals.[27]

An additional aspect to consider is the need for clear clinical supervision strategies regarding the use of AI by students and physicians in training. A recent study shows how the generational gap in mastering AI tools can create tensions in the teacher–student relationship and increase the risk of inappropriate use if adequate supervision frameworks are not in place.[27] Acquiring competencies in artificial intelligence is essential for the training of future physicians, particularly regarding ensuring the safe use of these tools. As noted by Ma et al. (2024), having a solid foundation in AI is necessary to assess risks and to understand phenomena such as "hallucinations." Incorporating such strategies into Spanish programs would not only facilitate a safer and more ethical adoption of AI but would also strengthen the role of educators as guides in the development of clinical judgment in an environment increasingly mediated by intelligent technologies.[4]

This study has several limitations that should be considered when interpreting the results. First, the analysis was based solely on publicly available information, which may not fully capture internal initiatives or unpublished development plans. Second, the categorization of AI content depended on the terminology used in official documents, which may have resulted in inconsistent classifications.

Future research should include direct surveys of universities and faculty to gain a deeper understanding of current and planned initiatives. Longitudinal studies would also be valuable to monitor progress in AI integration over time. Looking ahead, in addition to surveys and longitudinal studies, it would be advisable to establish national and international monitoring indicators to enable comparisons of the evolution of AI integration in curricula. Collaboration among faculties, scientific societies, and international organizations such as AMEE could facilitate the creation of global standards and promote translational research in digital education.[28]

In conclusion, this study provides the first systematic assessment of the state of artificial intelligence integration in undergraduate medical education in Spain, revealing a landscape that requires immediate attention and coordinated action. The findings show a delay in the integration of AI into medical curricula in Spain, underscoring the need for updated educational policies and the modernization of curricula to ensure that future healthcare professionals acquire competencies aligned with technological advances. The limited presence of AI training in the medical degree may place Spanish physicians at a disadvantage compared with their European counterparts, increasing the risk of errors due to misuse or excessive reliance on automated systems.

**Authors' contributions**

JT conceived and designed the study. AEJ and KPP curated and analysed the data, and conducted the investigation and methodology. AEJ and KPP created the visualisations and wrote the original draft. BPC and JT supervised and validated the study. JM, JC, FC, JBC, MRC, and JT critically reviewed and edited the manuscript. AEJ, KPP, BPC, and JT accessed and verified the underlying data. All authors had full access to all the data in the study, approved the final version of the manuscript, and accept responsibility for the decision to submit for publication.

**Acknowledgments**

AI use declaration: this manuscript made use of AI-assisted tools, including ChatGPT (GPT-4/GPT-5), Gemini 2.5 Pro, and Manus, to support information retrieval, analysis, drafting, and language revision. All outputs were generated under the supervision of the authors, who conducted the final review and editing, and take full responsibility for the content of the manuscript.

Ethical approval and patient consent were not required because this study did not include individual-level human data.

**Figures and tables**

Figure 1. Territorial Distribution of Universities with AI Courses in the Medical Degree Program in Spain

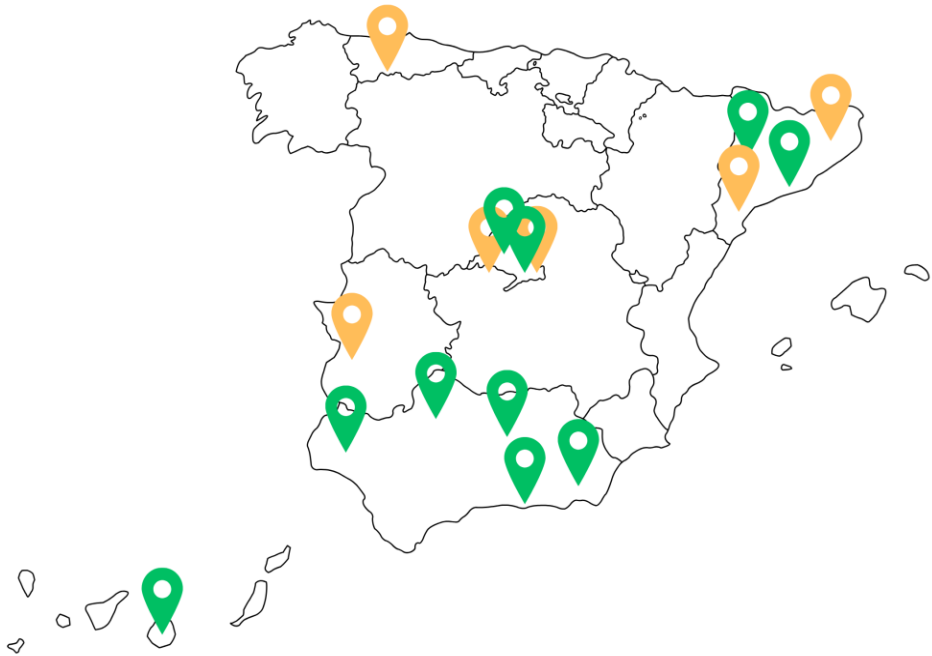

**52 universities with a Medical Degree Program
10 with Artificial Intelligence (AI) Courses**

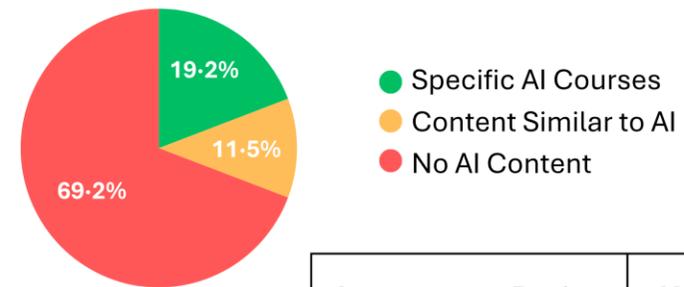

- Specific AI Courses
- Content Similar to AI
- No AI Content

19·2%
11·5%
69·2%

| Autonomous Region | AI Rate |
|---|---|
| Andalucía | 55·5% |
| Cataluña | 33·3% |
| Comunidad de Madrid | 25% |
| Islas Canarias | 20% |

**Only 1·17% of bachelor's credits (ECTS) are dedicated to AI Courses, which are primarily elective**

**Table 1: Consultation on Medical Degrees Offered in Spain: Register of Universities, Centers and Degrees (Registro de Universidades, Centros y Títulos - RUCT)**

| Search Field | Search Selection |
|---|---|
| University | All |
| Degree Name | Medicin |
| Academic Level | Bachelor's Degree (grado) |
| Field | Health Sciences |
| Area | Medicine and Dentistry |
| Status | All |
| Historical Search | No |

**Table 2: Universities with Specific AI Courses in the Medical Degree Program**

| University | Type | Region | Course | Status | ECTS | Level | Dept | Gen AI Inclusion |
|---|---|---|---|---|---|---|---|---|
| Autonomous University of Barcelona (UAB) | Public | Catalonia | Artificial Intelligence and Health (Inteligencia Artificial y Salud) | Elective | 3 | 3rd | Surgery | NO |
| Autonomous University of Madrid (UAM) | Public | Madrid | Artificial Intelligence (AI) in Biomedicine (Inteligencia Artificial (IA) en Biomedicina) | Elective/ Transversal | .. | .. | .. | .. |
| Complutense University of Madrid (UCM) | Public | Madrid | Practical Application of Generative Artificial Intelligence (Aplicación práctica de la inteligencia artificial generativa) | Elective | 3 | 1st, 2nd | Radiology, Rehabilitation, and Physiotherapy | Yes (Aplicación práctica de la IA-Gen) |
|  |  |  | Big Data and Artificial Intelligence in Medicine (Big Data e Inteligencia Artificial en Medicina) | Elective | 3 | 3rd, 6th | Medicine |  |
| European University of Madrid (UEM) | Private | Madrid | AI-Assisted Clinical Decision-Making: Ethics and Humanization (Decisiones clínicas asistidas por IA: ética y humanización) | .. | 3 | .. | .. | .. |
| Fernando Pessoa University, Canary Islands (UFPC) | Private | Canary Islands | Big Data and Artificial Intelligence in Medicine (Big Data e Inteligencia Artificial en Medicina) | Elective | 3 | .. | .. | .. |
| University of Huelva (UHU) | Public | Andalusia | Application of Big Data and Artificial Intelligence in Medicine (Aplicación de Big Data e Inteligencia Artificial en Medicina) | Elective | 3 | 4th | .. | .. |
| Loyola Andalusia University (ULA) | Private | Andalusia | Artificial Intelligence and Healthcare (Inteligencia Artificial y Atención Sanitaria) | Elective | 3 | 4th | .. | .. |

| University | Type | Region | Course | Status | ECTS | Level | Dept | Gen AI Inclusion |
|---|---|---|---|---|---|---|---|---|
| University of Lleida (UDL) | Public | Catalonia | Artificial Intelligence in Medicine (Inteligencia Artificial en Medicina) | Elective | 3 | 1st | Basic Medical Sciences | NO |
| Camilo José Cela University (UCJC) | Private | Madrid | Fundamentals of Artificial Intelligence in Healthcare (Fundamentos de la Inteligencia Artificial en atención médica) | Elective | 3 | 3rd | | YES |
| University of Córdoba (UCO) | Public | Andalusia | Documentation and Artificial Intelligence in Medicine. History of Medicine (Documentación e Inteligencia Artificial en Medicina. Historia de la Medicina) | Basic | 6 | 1st | Medical and Surgical Sciences | NO |
| University of Almería (UAL) | Public | Andalusia | Medical Informatics (Informática Médica) | Elective | 3 | 4th | Informatics | NO |
| University of Jaén (UJA) | Public | Andalusia | New Clinical and Biomedical Information Technologies (Nuevas tecnologías de la información clínica y biomédica) | Compulsory | 3 | 2nd | Microbiology | NO |

**Table 3. Universities with courses categorized as "Similar to IA"**

| University | Type | Region | Courses | Status | ECTS | Level | Dept. | AI Content Detail |
|---|---|---|---|---|---|---|---|---|
| University of Vic - Central University of Catalonia (UVic UCC) | Private | Catalonia | Information and Communication Technologies and Health (Tecnologías de la Información y la Comunicación y Salud) | Elective | 5 | $1^{st}, 2^{nd}$ | .. | It mentions AI in the objectives of existing courses and the university has a specific AI professorship. |
| Alfonso X El Sabio University (UAX) | Private | Madrid | Digital Health (Salud digital) | Elective | 6 | $6^{th}$ | .. | Among the contents detailed are: Artificial Intelligence (AI) in Value-based Healthcare; AI-based decision support tools; Which data are relevant in the healthcare setting and how to acquire, prepare, and store them; Data Mining; Natural Language Processing; Machine Learning, Success stories of Intelligent Systems such as IBM Watson: challenges, barriers, and facilitators |
| University of Extremadura (UNEX) | Public | Extremadura | Applied Medical Informatics (Informática Médica Aplicada) | Elective | 6 | $5^{th}$ | Department of Computer Systems and Telematics Engineering. Languages and Computer Systems | • Fundamentals of Informatics. Introduction to Medical Informatics.<br>• Search and management of health information.<br>• Communication and dissemination of health information.<br>• Health Information Systems. |
| Rovira i Virgili University (URV) | Public | Catalonia | New Technologies and Data Management (Nuevas Tecnologías y Manejo de Datos) | Compulsory | 3 | $3^{rd}$ | .. | Management and Analysis of Health Data with Advanced Technological Components |
| University of Oviedo (UNIOVI) | Public | Asturias | Pathological Anatomy (Anatomía Patológica)<br><br>Fundamentals of Surgery (Fundamentos de Cirugía) | .. | .. | .. | .. | Artificial intelligence is mentioned in the description of the course of Anatomical Pathology and during Fundamentals of Surgery in Topic 8. New challenges and future of surgery: Session 31. Telesurgery: robotics, virtual reality, and artificial intelligence |

| King Juan Carlos University (URJC) | Public | Madrid | Introduction to Medicine: Medical Information and Documentation (Introducción a la medicina: información y documentación medica) | Compulsory | 6 | 1st | Medical Specialties and Public Health | Module I. Information Management and Clinical-Research Questions (highlighting Unit 7: Fundamentals of Evidence-Based Medicine.). Formulating Problems/Questions: From the patient to the clinician-researcher. DEPTh Model. Questions and levels of training from the patient to the researcher. AI-based tools for EBM (Evidence-Based Medicine), specifically Generative Artificial Intelligence (GPTs, ChatGPT, Ellicit, others) for reading, analysis, and writing. |

**Table 4. Distribution of AI Courses by type: specific AI, Content Similar to AI, or without AI Content**

| University Type | Spanish Universities | Specific AI Course | Content Similar to AI | No AI Content |
|---|---|---|---|---|
| Public | 36 | 7 (19·4%) | 4 (11·1%) | 25 (69·4%) |
| Private | 16 | 3 (18·8%) | 2 (12·5%) | 11 (68·8%) |
| Total | 52 | 10 (19·2%) | 6 (11·5 %) | 36 (69·2%) |

**Supplementary Material**

**Table 1. List of Spanish universities offering a Medicine degree analyzed**

| University with a Medical Degree Program | Autonomous Region | Type | Web link | IA Course |
|---|---|---|---|---|
| Autonomous University of Barcelona (UAB) | Catalonia | Public | https://www.uab.cat/ | Yes |
| Complutense University of Madrid (UCM) | Madrid | Public | https://www.ucm.es/ | Yes |
| University of Vic - Central University of Catalonia (UVic UCC) | Catalonia | Private | https://www.uvic.cat/es | Similar to AI |
| Fernando Pessoa University, Canary Islands (UFPC) | Canary Islands | Private | https://ufpcanarias.es/ | Yes |
| University of Huelva (UHU) | Andalusia | Public | https://www.uhu.es/ | Yes |
| Autonomous University of Madrid (UAM) | Madrid | Public | https://www.uam.es/ | Yes |
| University of Barcelona (UB) | Catalonia | Public | https://web.ub.edu/ | No |
| Alfonso X El Sabio University (UAX) | Madrid | Private | https://www.uax.com/ | Similar to AI |
| University of Zaragoza (UNIZAR) | Aragon | Public | https://www.unizar.es/ | No |
| Pompeu Fabra University (UPF) | Catalonia | Public | https://www.upf.edu/ | No |
| University of Valencia (UV) | Valencian Community | Public | https://www.uv.es/ | No |
| University of Seville (US) | Andalusia | Public | https://www.us.es/ | No |
| University of Granada (UGR) | Andalusia | Public | https://www.ugr.es/ | No |
| University of Salamanca (USAL) | Castile and León | Public | https://www.usal.es/ | No |
| University of Murcia (UMU) | Region of Murcia | Public | https://www.um.es/ | No |
| University of the Basque Country (UPV/EHU) | Basque Country | Public | https://www.ehu.eus/ | No |
| University of Navarra (UNAV) | Navarre | Private | https://www.unav.edu/ | No |
| CEU San Pablo University | Madrid | Private | https://www.uspceu.com/ | No |
| European University of Madrid (UEM) | Madrid | Private | https://universidadeuropea.com/ | Yes |
| Francisco de Vitoria University (UFV) | Madrid | Private | https://www.ufv.es/ | No |
| Catholic University of Valencia San Vicente Mártir (UCV) | Valencian Community | Private | https://www.ucv.es/ | No |
| Catholic University of Murcia (UCAM) | Region of Murcia | Private | https://www.ucam.edu/ | No |
| International University of Catalonia (UIC) | Catalonia | Private | https://www.uic.es/ | No |
| University of Castile-La Mancha (UCLM) | Castile-La Mancha | Public | https://www.uclm.es/ | No |

| University of Extremadura (UNEX) | Extremadura | Public | https://www.unex.es/ | Similar to AI |
| --- | --- | --- | --- | --- |
| University of Oviedo (UNIOVI) | Asturias | Public | https://www.uniovi.es/ | Simiar to AI |
| University of Loyola Andalusia (ULA) | Andalusia | Private | https://www.uloyola.es/ | Yes |
| University of Santiago de Compostela (USC) | Galicia | Public | https://www.usc.gal/ | No |
| University of Málaga (UMA) | Andalusia | Public | https://www.uma.es/ | No |
| University of Cádiz (UCA) | Andalusia | Public | https://www.uco.es/ | Yes |
| University of Cantabria (UCN) | Cantabria | Public | https://web.unican.es/ | No |
| University of Alcalá (UAH) | Madrid | Public | https://www.uah.es/ | No |
| University of La Laguna (ULL) | Canary Islands | Public | https://www.ull.es/ | No |
| University of the Balearic Islands (UIB) | Balearic Islands | Public | https://www.uib.eu/ | No |
| University of Lleida (UDL) | Catalonia | Public | https://www.udl.cat/ | Yes |
| University of Girona (UDG) | Catalonia | Public | https://www.udg.edu/ | No |
| Jaume I Castellón University (UJI) | Valencian Community | Public | https://www.uji.es/ | No |
| Miguel Hernández University of Alicante (UMH) | Valencian Community | Public | https://www.umh.es/c | No |
| King Juan Carlos University (URJC) | Madrid | Public | https://www.urjc.es/ | Similar to AI |
| University of Valladolid (UVA) | Castile and León | Public | https://med.uva.es/ | No |
| CEU Cardenal Herrera Castellón University (UCHCEU) | Valencian Community | Private | https://www.uchceu.es/ | No |
| Antonio de Nebrija University (UAN) | Madrid | Private | https://www.nebrija.com/ | No |
| Camilo José Cela University (UCJC) | Madrid | Private | https://www.ucjc.edu/ | Yes |
| University of Alicante (UA) | Valencian Community | Public | https://web.ua.es/ | No |
| University of Almería (UAL) | Andalusia | Public | https://www.ual.es/ | Yes |
| University of Cádiz (UCA) | Andalusia | Public | https://medicina.uca.es/ | No |
| University of Deusto (UD) | Basque Country | Private | https://www.deusto.es/ | No |
| University of Jaén (UJA) | Andalusia | Public | https://uvirtual.ujaen.es/ | Yes |
| University of Las Palmas de Gran Canaria (ULPGC) | Canary Islands | Public | https://www2.ulpgc.es/ | No |
| Public University of Navarra (UPNA) | Navarra | Public | https://www.unavarra.es/ | No |
| Rovira i Virgili University (URV) | Catalonia | Public | https://www.urv.cat/ | Similar to AI |
| San Jorge University (USJ) | Aragon | Private | https://www.usj.es/ | No |

**Table 2. Content of the specific AI courses in the Medicine degree program**

| University | Course | Status | ECTS | Level | Content |
|---|---|---|---|---|---|
| Autonomous University of Barcelona (UAB) | Artificial Intelligence and Health (Inteligencia Artificial y Salud) | Elective | 3 | 3rd | - Introducción a la inteligencia artificial y el aprendizaje automático.<br>- Medicina y Cirugía basada en la evidencia. Normalización lingüística. Motores de búsqueda.<br>- Entorno smart city. Smart Health. Hospital Líquido. El papel del médico en un entorno Smart Health.<br>- Biometría del medio y Big Data. Internet of Things. App y Telemetría.<br>- Computación neuromórfica. Aprendizaje profundo. Modelos predictivos supervisados y no supervisados.<br>- El cerebro médico global.<br>- Robótica aplicada al ámbito asistencial.<br>- Bioética del Aprendizaje automático. Ética algorítmica. |
| Complutense University of Madrid (UCM) | Practical application of generative artificial Intelligence (Aplicación práctica de la inteligencia artificial generativa) | Elective | 3 | 1st, 2nd | **Practical application of generative artificial intelligence.**<br>- Generative AI in Research: Introduction to generative AI and its use in medical research.<br>- Generative AI in Education and Training: The use of generative AI in medical education and training.<br>- Generative AI and Clinical Decision-Making: Application of generative AI in clinical decision-making.<br>- AI and Innovation: Exploring the role of AI in medical innovation. |
| | Big Data and Artificial Intelligence in Medicine (Big Data e Inteligencia Artificial en Medicina) | Elective | 3 | 3rd, 6th | **Big Data and Artificial Intelligence in Medicine.**<br>- Does AI add value to clinical medicine?<br>- The future of medical language: natural language processing and new options in medical information management.<br>- The use of AI in medical image management.<br>- AI-based clinical research versus "orthodox" clinical research. Secondary analysis of medical databases.<br>- Introduction to the algorithms used in Machine Learning and Artificial Intelligence.<br>- New information technologies and Big Data: an update.<br>- Introduction to Machine Learning. Crowdsourcing of health data. |
| | | | | | - AI in the prediction of clinical syndromes. Is it better than an experienced doctor?<br>- AI as "Point of Care" for clinical decision-making?<br>- The quality of data in big data.<br>- AI in therapeutic decision-making.<br>- AI in the pre-hospital setting.<br>- Can it be implemented throughout the hospital?<br>- AI in Intensive Care Medicine. Can it change the quality of care?<br>- Introduction to Practical Sessions. |
| University of Lleida (UdL) | Artificial Intelligence in Medicine (Inteligencia Artificial en Medicina) | Elective | 3 | 1st | - Introduction to Artificial Intelligence in Medicine<br>- Machine Learning Fundamentals<br>- Data Acquisition and Preprocessing<br>- Medical Image Analysis<br>- Clinical Decision Support Systems<br>- Natural Language Processing (NLP) in Medicine<br>- Ethical and Legal Considerations<br>- Future Directions and Emerging Trends<br>- Practical Applications and Case Studies<br>- Discussion and Future Implications |
| Camilo José Cela University (UCJC) | Fundamentals of Artificial Intelligence in Healthcare (Fundamentos de la Inteligencia Artificial en atención médica) | Elective | 3 | 3rd | **Theoretical Content:**<br><br>- Introduction to AI in Medicine. History and evolution of AI in the medical field. Foundational concepts.<br>- Types of AI algorithms. Machine Learning vs. Deep Learning. Basic applications in medicine.<br>- Neural networks in medicine. Main architectures. Use cases in diagnosis.<br>- AI-assisted diagnosis. Clinical decision support systems. Validation of results.<br>- AI in medical imaging. Processing of radiological images. Pathology detection.<br>- Digital pathology. Automated sample analysis. Classification systems.<br>- AI in clinical prediction. Predictive models. Risk assessment.<br>- Personalized medicine and AI. Genomics and AI. Personalized treatments.<br>- Natural Language Processing in medicine. Analysis of medical records. Extraction of medical information.<br>- Large Language Models in medicine. GPT and other models. Clinical applications.<br>- Ethics in medical AI. Privacy and security. Professional responsibility.<br>- Regulation of AI in medicine. Current legal framework. System certification.<br>- Future of AI in medicine. Emerging trends. Challenges and opportunities. |
| | | | | | - Integration of AI into clinical practice. Success stories. Lessons learned.<br><br>**Practical Content:**<br><br>- Introduction to AI Tools in Medicine. Setting up the work environment.<br>- Basic Prompt Engineering. Fundamental principles. Practical exercises.<br>- Advanced Prompt Engineering. Optimization techniques. Practical cases.<br>- ChatGPT in Medicine I. Basic configuration and use. Basic medical consultations.<br>- ChatGPT in Medicine II. Creating specialized prompts. Analysis of clinical cases.<br>- Custom GPTs for Medicine. Development of specialized assistants. Configuration and testing.<br>- Claude in Medicine I. Specific features. Comparison with other models.<br>- Claude in Medicine II. Development of specific applications. Practical cases.<br>- Perplexity for Research I. Search and analysis of medical literature. Research techniques.<br>- Perplexity for Research II. Assisted scientific writing. Bibliographic review.<br>- Computer Vision in Medicine I. Analysis of medical images. Tumor detection.<br>- Computer Vision in Medicine II. Image segmentation. Automatic classification.<br>- Development of an Assistant for Scientific Articles I. Design and planning. Initial implementation.<br>- Development of an Assistant for Scientific Articles II. Refinement and testing. Final presentation. |
| University of Córdoba (UCO) | Documentation and Artificial Intelligence in Medicine. History of Medicine (Documentación e Inteligencia Artificial en Medicina. Historia de la Medicina) | Basic | 6 | 1st | **THEORETICAL CONTENTS:**<br>**BLOCK I. INTRODUCTION TO THE HISTORY OF MEDICINE.**<br>- Diseases and society.<br>- Medical systems.<br>- Modern scientific medicine from the Renaissance to the beginning of the 19th century.<br>- Contemporary scientific medicine from the 19th to the 21st century<br><br>**BLOCK II: MEDICAL DOCUMENTATION**<br>**Sub-block I. Scientific documentation.**<br>- Scientific documentation.<br>- Primary sources of information.<br>- Secondary sources of information.<br>- Tertiary sources of information. |
| | | | | | • Analysis of scientific literature<br><br>**Sub-block II. Clinical documentation.**<br>• Clinical documentation: introduction, concept, and typology.<br>• Admission and Clinical Documentation Services.<br>• Clinical History: historical evolution, concepts, and Minimum Basic Data Set (MBDS).<br>• Medical terminology.<br>• Ethical use of information<br><br>**BLOQUE III. INTELIGENCIA ARTIFICIAL EN MEDICINA**<br>• Introduction to IA y Big Data.<br>• Introduction to Machine Learning.<br>• Chatbot.<br>• Natural Language Processing.<br><br>**PRACTICAL CONTENTS**<br>• Primary sources of information: Scientific journals.<br>• Secondary sources of information: Search for biomedical information.<br>• Tertiary sources of information: Evidence-Based Medicine (EBM).<br>• Clinical documentation.<br>• Medical terminology.<br>• AI in Medicine (I).<br>• AI in Medicine (II). |
| University of Almería (UAL) | Medical Informatics (Informática Médica) | Elective | 3 | 4th | **Block I:**<br>• Introduction to Artificial Intelligence: Contextualizes the development and evolution of AI, as well as its current healthcare applications.<br>• Image Analysis: Explores computer vision techniques, which are key in specialties like radiology, dermatology, and digital pathology.<br>• Medical AI Practice: Promotes active learning through the use of AI tools in real or simulated cases.<br>**Block II:**<br>• Natural Language Processing (NLP): Teaches how to interpret and generate clinical language using NLP models, improving computer-assisted decision-making.<br>• Evaluation, Challenges, and Ethical Issues of AI: Fosters reflection on the reliability, biases, explainability, and legal aspects of medical AI. |
| | | | | | • AutoML and Practice Presentation: Introduces the automation of machine learning, allowing students to build models without needing extensive programming experience, and to share their developments<br><br>**Block III:** Focuses on the complete workflow related to digital medical imaging, from its acquisition to its visualization, storage, and transmission<br>**Block IV**: Addresses the application of automation, control, and robotics techniques to the medical environment, with special emphasis on therapeutic and diagnostic support devices. |
| University of Jaén (UJA) | New Clinical and Biomedical Information Technologies (Nuevas tecnologías de la información clínica y biomédica) | Compulsory | 3 | 2nd | Second Part: Artificial Intelligence and Big Data Processing in Biomedicine<br>• Machine Learning in Medical Imaging<br>• Health Intelligence<br>• Artificial Intelligence in Bioinformatics: Development of an Automated Methodology for Predicting the Protein Residue Contact Map<br>• Deep Learning in Biomedical Image Analysis<br>• Automatic Detection of Lesions with Three-Dimensional Convolutional Neural Networks<br>• Biomedical Image Segmentation for Oncology and Precision Radiation<br>• Towards Large-Scale Histopathological Image Analysis using Deep Learning |
| Fernando Pessoa University, Canary Islands (UFPC) | Big Data and Artificial Intelligence in Medicine | Elective | 3 | .. | .. |
| University of Huelva (UHU) | Application of Big Data and Artificial Intelligence in Medicine | Elective | 3 | .. | .. |
| Loyola Andalusia University (ULA) | Artificial Intelligence and Healthcare | Elective | 3 | 4th | .. |

There is no information available for Fernando Pessoa University, Canary Islands (UFPC), the University of Huelva (UHU), or Loyola Andalusia University (ULA) because no details are available regarding the content of their courses